\documentclass[final,3p,times,twocolumn]{elsarticle3}
\usepackage{amssymb}
\usepackage{graphicx}
\usepackage{bm}

\journal{Studies in History and Philosophy of Modern Physics, October 1, 2016}

\begin{document}

\begin{frontmatter}



\title{Quantum jumps, superpositions, and the continuous evolution of quantum states}


%%\author{Removed as required by publisher for double blind review}
%%\ead{Removed as required by publisher for double blind review}

\author{Rainer Dick}
\ead{rainer.dick@usask.ca}
%%\author{A.B. Other}

%%\address{Removed as required by publisher for double blind review}

\address{Department of Physics and Engineering Physics, 
University of Saskatchewan, Saskatoon, Canada SK S7N 5E2\\
}

\begin{abstract}
The apparent dichotomy between quantum jumps on the one hand,
and continuous time evolution according to wave equations on the other hand,
provided a challenge to Bohr's proposal of quantum jumps in atoms.
 Furthermore, Schr\"odinger's time-dependent equation also seemed to
require a modification of the explanation for the origin of line
spectra due to the apparent possibility of superpositions of energy
eigenstates for different energy levels.
Indeed, Schr\"odinger himself proposed a quantum beat mechanism for the generation
of discrete line spectra from superpositions of eigenstates with different energies.

However, these issues between old quantum theory and Schr\"odinger's
wave mechanics were correctly resolved only after the development 
and full implementation of photon quantization.
The second quantized scattering matrix formalism reconciles quantum jumps with 
continuous time evolution through the identification of quantum jumps with
transitions between different sectors of Fock space. The
continuous evolution of quantum states is then recognized as a sum over continually 
evolving jump amplitudes between different sectors in Fock space.

In today's terminology, this suggests that linear combinations of scattering matrix 
elements are epistemic sums over ontic states.
Insights from the resolution of the dichotomy between quantum jumps and continuous
time evolution therefore hold important lessons for modern research both on 
interpretations of quantum mechanics and on the foundations of quantum computing.
They demonstrate that discussions of interpretations of quantum theory necessarily
need to take into account field quantization. They also demonstrate the limitations
of the role of wave equations in quantum theory, and caution us that superpositions of 
quantum states for the formation of qubits may be more limited than usually expected.
\end{abstract}


\end{frontmatter}

%%%%%%%%%%%%%%%%%%%%%%%%%%%%%%%%%%%%%%%%%%%%%%%%%%%%%%%%%%%%%%%%%%%%%%%%
\section{Introduction}
\label{sec:intro}
%%%%%%%%%%%%%%%%%%%%%%%%%%%%%%%%%%%%%%%%%%%%%%%%%%%%%%%%%%%%%%%%%%%%%%%%

The question for the interpretation of quantum states remains an unsolved
and still controversial issue ever since Schr\"odinger introduced his 
evolution equation for quantum states as a function of time, which also 
opened the door for deterministic interpretations \cite{erwin}. We know 
how to combine causal evolution of quantum states with Born's prescription 
to relate quantum states to observables, and the most precise physical 
constants have been calculated using this formalism.
However, the final step of projecting out one particular part of a continually
evolved quantum state, purportedly in response to a question that we ask
through an experimental probe, {\it prima facie} seems to introduce a somewhat 
mysterious discontinuous dynamical aspect into quantum mechanics.
This puzzling aspect of quantum mechanics has been investigated from many
angles and in many experimental settings, most notably in particle diffraction
and Stern-Gerlach type experiments. The atomic emission problem does
not enjoy the same level of prominence, although it has played an important
role in Schr\"odinger's concerns about the foundations of quantum mechanics,
and there are aspects in the explanation of the emergence of spectral lines
in spontaneous emission which are still challenging from the point of view
of the Copenhagen interpretation\footnote{The notion of
a Copenhagen interpretation of quantum mechanics can mean different things
for different experts in the field. Here ``Copenhagen interpretation''
refers to Born's interpretation of projections of ontic quantum states
onto eigenstates of hermitian operators as probability amplitudes
(or amplitudes for probability densities), combined with the assertion
that the resulting probabilities measure the likelihood that individual
quantum systems (e.g. single particles or few particle systems)
collapse into the eigenstate if stimulated through a corresponding observation.}.

Many of the more problematic aspects of the Copenhagen interpretation can be 
avoided or resolved in epistemic interpretations of quantum states as records 
of collective or individual knowledge of observers performing an 
experiment \cite{fuchs,qb1,ferrero,spekkens}. There is no need for apparent
spontaneous or even retroactive collapse of ontic quantum states in epistemic 
interpretations, and Schr\"odinger cat states do not refer to ontological 
properties of a system. Epistemic interpretations therefore challenge the realist 
or ontic interpretation of quantum states as records of complete, objective 
information of ontological properties of physical systems. However, it is likely 
a fair assessment that ``quantum practitioners'', while mostly remaining agnostic 
with respect of the difficult philosophical aspects of quantum mechanics, tacitly 
use a realist interpretation of quantum states. Furthermore, epistemic 
interpretations have recently been confronted with numerous counterarguments on 
the basis of quantitative and observational analyses of entangled states
\cite{realpsia,realpsi0,realpsi1,realpsi2,realpsi3,realpsi4,realpsi5,realpsi6},
which in turn have been challenged through full locality 
in the QBism framework \cite{qb2}. 

Another interesting way to avoid puzzling aspects of the traditional Copenhagen 
interpretation is the statistical ensemble interpretation of quantum 
mechanics \cite{ballentine1,ballentine2}. This emphasizes that the Born rule 
relates predictions from single particle wavefunctions to experiments with many 
particles, but that we cannot infer predictions for single particles from the 
wavefunction. In particular, there is no need for quantum state collapse if 
single particle wavefunctions cannot be directly associated with the behavior 
of individual particles. This proposal in itself does not necessarily favor either 
an ontic or an epistemic interpretation of quantum states. The flipside of the 
statistical ensemble interpretation (in my understanding) is that we would have 
to resign ourselves to the fact that we do not have access to any single particle 
dynamics, or otherwise assume that quantum mechanics is only a statistical 
approximation to a yet to be discovered single particle dynamics. 

The amount of intellectual effort and capacity spent on the interpretation of 
quantum mechanics, and the fact that the discussion has intensified rather than 
subsided over the past 90 years, provide testimony to the difficulty of the 
underlying interpretational problems. The present paper tries to shed light on 
these questions from the perspective of the theory of photon emission and 
absorption by atoms. The underlying calculations are well-known standard 
applications of quantum optics. However, I would like to argue that paying 
attention not only to the overall results of the calculations, but to the
derivation and to particular details of the results, helps to put the 
quest for a universally acceptable interpretation of quantum states into 
even stronger focus. I hope to convince the reader that revisiting the atomic 
emission problem as a case study for what everyday practical calculations reveal 
about quantum systems, yields interesting insights into the nature of quantum 
states, including the possibility of epistemic combinations of ontic states 
in Fock space.

 For example, we will see that the scattering matrix elements for optical 
emission and absorption both for single photons and for coherent states
show that the formalism of calculating quantum mechanical observables 
assigns special significance to energy eigenstates even if the observables
do not include the energy of any component of a system. More specifically,
quantum jumps between energy eigenstates seem to be an integral part of the 
response of every quantum system no matter how we probe the system.
How can that be if all unitarily equivalent bases of quantum states are
physically equivalent? What makes energy eigenstates a preferred part of
the response of a system even if we are not probing for energies?
On the other hand, how is it possible to reconcile quantum jumps 
and continuous evolution of quantum states?

Besides illuminating the problem of the interpretation of quantum states and 
the measurement process from a different angle, reviewing quantum optical 
calculations with a critical view on what they really tell us is also of 
renewed practical interest. Demonstrations of single photon emissions from 
quantum dots \cite{1pem1,1pem2,1pem3} and the prospects to use these 
devices for quantum cryptography and quantum information processing force us 
to sharpen our understanding of the basic quantum optical processes of photon 
emission, absorption, and observation. Moreover, progress in attosecond 
spectroscopy \cite{atto1,atto2} now affords observations of electrons with 
atomic time scale resolutions \cite{atom1,atom2,atom3,atom4,atom5}.
The non-perturbative nature of the strong electromagnetic fields involved in
these experiments led to theoretical analyses of the observations through 
continuous evolution of atomic wavefunctions between different energy levels
within the confines of semi-classical approximations. This is born out of
practical calculational necessity and should be justified on the basis of high 
photon numbers in the laser pulses. However, semi-classical calculations for
optical responses of small numbers of atoms, described through continuously
evolving wavefunctions, are also at the heart of de Broglie-Bohm or 
Schr\"odinger type interpretations of quantum dynamics\footnote{I will denote 
theories or interpretations which assume that formulations of quantum dynamics 
should be possible in terms of continually evolving wavefunctions without any 
need for quantum jumps as ``de Broglie-Schr\"odinger type''. This is motivated 
by de Broglie's early proposal of de Broglie-Bohm type pilot wave theories for 
continuous particle trajectories and by Schr\"odinger's early interpretations 
of redundancy of wave-particle duality in a theory of continually evolving 
wavefunctions. Indeed, Schr\"odinger's early work on quantum mechanical wave 
equations was influenced by de Broglie's ideas, but the vagueness of his early 
statements on the interpretation of the wavefunction seem to indicate a readiness 
to move beyond a matter wave picture and could even be interpreted as early hints 
at a statistical interpretation, see in particular the paragraph on pp. 134-135 
in Ref. \cite{erwin}. However, later in life Schr\"odinger clearly embraced de 
Broglie's ideas on matter waves \cite{erwin2,erwin3}, although he 
did not consider them as pilot waves for particles. The early developments of 
de Broglie-Bohm theories and of Schr\"odinger's interpretation are beautifully 
reviewed in \cite{Holland} and \cite{perovic}, respectively.}, which assume that 
continually evolving wavefunctions can provide a complete description of quantum 
evolution without a need for quantum jumps \cite{erwin2,erwin3}. 
This is in contrast to the well-established second quantized scattering 
matrix formalism, which yields continually evolving probability amplitudes for 
quantum jumps.

Therefore, a particular topic that we will reconsider for its relevance to 
single photon generation and detection, and for its potential to afford us a 
better understanding of the foundations of quantum dynamics and the meaning of 
quantum states, is the emergence of line spectra from atomic states. We will see 
that the issues of quantum jumps, photon quantization, scattering matrices for 
photon emission or absorption, and interpretation of quantum states in atom-photon 
systems are all intertwined. The second quantized scattering matrix will play a 
central role in our analysis, but we need to develop a brief historic perspective 
on Schr\"odinger's wave ontology in relation to line spectra to fully appreciate 
what the scattering matrix has to tell us. This will be done in Section \ref{sec:erwin}.
We will see in particular that combinations of the established wave equation for
non-relativistic particles, \textit{viz.} Schr\"odinger's equation, with Maxwell's 
equations for radiation cannot support Schr\"odinger's proposal of continuous
(i.e. without quantum jumps) generation of radiation from atoms. This observation
reconfirms the central role of the second quantized scattering matrix formalism for
the description of emission from atoms. 

The emission rates and absorption cross sections are therefore revisited for 
mono\-chro\-matic photon states in Section \ref{sec:mono}, and for coherent photon 
states in Section \ref{sec:coherent}. Section \ref{sec:conc} discusses our conclusions.

Details of the standard derivations and results for photon emission and absorption 
in quantum theory are particularly relevant for the present investigation.
They will tell us that quantum jumps are unavoidable for the formation of line
spectra. The scattering matrices for coherent photon states will also
demonstrate that collapse into energy eigenstates is an inevitable phenomenon
in quantum optics even in cases where we do not observe the energy of any 
component of the system at any time. This is clearly worrisome for traditional
proposals of observer or apparatus induced state collapse. On the other hand,
there is also no space here for any critical participation of the environment
in the formation of line spectra. The results rather indicate a mixed ontic
and epistemic interpretation of quantum state evolution as encoded in scattering
matrices: Scattering matrices generically evolve ontic $N$-particle states
into epistemic sums spanning several sectors of Fock space, i.e. scattering
matrices describe epistemic sums over ontic states.

The phrases ``quantum jump'' and ``quantum leap'' are usually interchangeable
synonyms in physics. Here the word ``jump'' will be used to designate sudden,
discontinuous transitions between ontic quantum states with the underlying 
assumption that the transition happens between the states without any necessary 
involvement of an observer. On the other hand, the word ``leap'' will be 
associated with a discontinuous reassignment of a quantum state to a system 
by an observer, as a consequence of a change of knowledge about the system.
This is not a clear call in many instances, and in cases where neither an 
ontic nor an epistemic interpretation appears warranted, the word ``jump'' 
will be used as the default option.

%%%%%%%%%%%%%%%%%%%%%%%%%%%%%%%%%%%%%%%%%%%%%%%%%%%%%%%%%%%%%%%%%%%%%%%%
\section{Schr\"odinger's explanation for line spectra}
\label{sec:erwin}
%%%%%%%%%%%%%%%%%%%%%%%%%%%%%%%%%%%%%%%%%%%%%%%%%%%%%%%%%%%%%%%%%%%%%%%%

It is prudent to revisit Schr\"odinger's proposal for formation of line 
spectra from continually evolving wavefunctions before taking the second 
quantized scattering matrix formalism and quantum jumps for granted.

If the time-independent Schr\"odinger equation did not have a 
time-dependent counterpart, the existence of discrete energy eigenstates 
in atoms would be sufficient to explain the observation of line spectra 
because we could only form superpositions of eigenstates within one energy 
level. The time-dependent Schr\"odinger equation invalidates this simple 
reasoning, and therefore the postulate that a single quantum system (like 
an atom) always responds with an eigenvalue of the observed 
quantity \cite{vonNeumann} is usually invoked to argue that observation of
emitted radiation yields observed values in terms of eigenvalues of atomic
Hamiltonians,
\begin{equation}\label{eq:jump1}
E_\gamma=\hbar ck=E_{n,\ell}-E_{n',\ell'}.
\end{equation}

The underlying problem is the superposition principle for solutions of the 
time-dependent Schr\"odinger equation. If hydrogen atoms can exist in a 
unitarily evolving superposition of bound energy 
eigenstates\footnote{We neglect a possible admixture of Coulomb waves for
simplicity. The purpose is to study the conceptual problem of line
spectra, for which linear combinations of bound states are sufficient.}
\begin{equation}\label{eq:super1}
|\{C\}(t)\rangle=\sum_{n,\ell,m_\ell}C_{n,\ell,m_\ell}|n,\ell,m_\ell\rangle
\exp(-\mathrm{i}E_{n,\ell}t/\hbar),
\end{equation}
with $\sum_{n,\ell,m_\ell}|C_{n,\ell,m_\ell}|^2=1$ and energy expectation value
\[
\langle E\rangle(\{C_{n,\ell,m_\ell}\})
=\sum_{n,\ell,m_\ell}|C_{n,\ell,m_\ell}|^2E_{n,\ell},
\]
why does a transition to a new state with expansion coefficients $C'_{n,\ell,m_\ell}$
and lower energy expectation value not lead to observation of photons of energy
$E_\gamma=\langle E\rangle(\{C_{n,\ell,m_\ell}\})-\langle E\rangle(\{C'_{n,\ell,m_\ell}\})$?
Stated differently: Why does emission or absorption of photons by atoms only 
involve transitions between energy eigenstates in agreement with Bohr's explanation
of atomic transitions in old quantum theory? 
Is it possible to demonstrate that unperturbed atoms should only exist in energy
eigenstates in spite of the fact that (\ref{eq:super1}) is a perfectly viable
solution of the time-dependent Schr\"odinger equation?
Or otherwise, is it possible to explain 
observation of line spectra \textit{without} inferring that the atomic transitions 
involve pure energy eigenstates as initial and final states of the emitting or 
absorbing atoms?

These questions about the role of unitarily evolving superpositions of energy
eigenstates arose as soon as Schr\"odinger published the time-dependent version of 
his wave equation \cite{erwin}, when superposition of atomic eigenstates for 
different energies instead of stationary Bohr orbits became an apparent possibility for 
the state of an atom. However, Schr\"odinger did not consider this as a potentially 
problematic challenge to the direct transition between Bohr orbits in old quantum 
theory, but rather as an opportunity to propose an alternative explanation for line 
spectra through the beats in the electric fields of atomic electric charge and current 
densities from the time-dependent interference terms in
\[
\varrho_e(\bm{x},t)=-\,e|\langle\bm{x}|\{C\}(t)\rangle|^2
\]
and
\begin{eqnarray}\nonumber
\bm{j}_e(\bm{x},t)&=&-\,\frac{e\hbar}{2\mathrm{i}m}[
\langle\{C\}(t)|\bm{x}\rangle\bm{\nabla}\langle\bm{x}|\{C\}(t)\rangle
\\ \label{eq:je}
&&-\,\bm{\nabla}\langle\{C\}(t)|\bm{x}\rangle\cdot\langle\bm{x}|\{C\}(t)\rangle],
\end{eqnarray}
see pp. 121 and 129-130 in Ref. \cite{erwin}. 

Indeed, the photon-matter interaction term in leading order in minimal coupling is
\begin{equation}\label{eq:H1b}
H_I=-\int\!d^3\bm{x}\,\bm{A}(\bm{x},t)\cdot\bm{j}_e(\bm{x},t),
\end{equation}
and this yields the leading order scattering matrix element for
photon absorption or emission,
\begin{equation}\label{eq:sfilast}
S_{\!fi}=\frac{\mathrm{i}}{\hbar}\int_{-\infty}^\infty\!dt
\int\!d^3\bm{x}\,\bm{A}(\bm{x},t)\cdot\bm{j}_e(\bm{x},t),
\end{equation}
where $\bm{A}(\bm{x},t)$ is the vector potential corresponding to the absorbed or 
emitted photon. The time integration in (\ref{eq:sfilast}) forces the Fourier 
components of $\bm{A}(\bm{x},t)$ to match the atomic 
transition frequencies in (\ref{eq:je}), whereas all the 
other modes suffer destructive interference. This yields the line spectra 
even for transitions between superpositions of atomic eigenstates and even if 
the photon states are not monochromatic. From a de Broglie-Schr\"odinger perspective
this looks like the absorption or emission of radiation arises as a consequence
of transient beats in the intra-atomic electric current. In stark contrast to the
traditional interpretation of the very same scattering matrix element (\ref{eq:sfilast}),
it would not arise as a consequence of a quantum jump from one energy eigenstate
to another energy eigenstate, but rather as a consequence of a temporary superposition
of both energy eigenstates which may arise as a consequence of a perfectly smooth
time evolution of the atomic state.

Schr\"odinger was mostly concerned with applications of his time-dependent wave 
equation to light scattering in Ref. \cite{erwin}, i.e. Schr\"odinger offered a 
qualitative explanation for emission and absorption of light through beats in the 
wavefunction, but he did not explicitly write down the coupling term (\ref{eq:H1b})
nor did he calculate corresponding emission and absorption probabilities.
Emission and absorption probabilities were calculated by Dirac in the semi-classical 
dipole approximation two months later \cite{dirac1a}. However, Dirac abstained from 
any speculation as to the origin of the transition, i.e. he did not address the
question whether he was calculating 
probabilities for quantum jumps in agreement with Bohr's proposal from old quantum 
theory, or whether he was dealing with consequences of quantum beats as Schr\"odinger 
had mused two months earlier. On p. 264 of his follow-up paper \cite{dirac1b}, which 
went beyond the semi-classical approximation by coupling a quantized electric field 
into the Schr\"odinger equation, Dirac does explicitly refer to quantum jumps. 
Dirac does not offer an explanation for his choice of interpretation, but there may 
be two reasons for his preference for quantum jumps: The actual calculation of 
emission and absorption coefficients through integration of the time-dependent 
Schr\"odinger equation does not require a superposition of energy eigenstates for 
the initial state, but naturally evolves even a pure energy eigenstate into a 
superposition of eigenstates. The intuitive (although not logically inevitable) 
conclusion from this is that quantum beats are not necessary for emission and 
absorption of radiation\footnote{The conclusion is not logically inevitable, because 
coupling to the electromagnetic field immediately evolves a pure energy eigenstate 
into a superposition of atomic eigenstates, which could be read to imply quantum 
beats in the sense of Schr\"odinger within the confines of semi-classical 
approximations.}. Furthermore, Schr\"odinger himself cautioned on p. 130 
of \cite{erwin} that a non-linear feedback mechanism would have to be added to his 
equation to terminate emission from quantum beats and explain how the system relaxes 
into the groundstate, and this may have stalled further pursuit of his idea. 
Schr\"odinger reiterated the idea of radiation from quantum beats in atoms much later 
in \cite{erwin2,erwin3}, but again only in a qualitative sense.

It is noteworthy that there is no need to invoke observers or probes in the quantum 
beat interpretation, nor is there any need to spontaneously collapse states to energy 
eigenstates at any time. In short, Schr\"odinger's proposal amounted to a trade off 
between actual electronic oscillations in atoms on the one hand versus some of the 
more problematic aspects of the emerging Copenhagen interpretation on the other hand. 
In this regard it is amusing to note that on p. 133 of his famous textbook on atomic 
physics \cite{born}, Born also proposed a quantum beat interpretation of the dipole 
matrix element $e\langle n',\ell',m'_\ell|\mathrm{\bf x}|n,\ell,m_\ell\rangle$ which appears
in scattering elements for photon emission.

However, insurmountable weaknesses of Schr\"odinger's proposal of radiation from 
quantum beats are the reliance on a semi-classical description of electromagnetic
 fields, and that he never developed it into a mathematical theory for emission line 
strengths. Schr\"odinger mentions the need for nonlinear coupling terms for wavefunctions 
in \cite{erwin3}, but he never writes down a coupled Schr\"odinger-Maxwell system
to see how far his ideas might carry. Indeed, he would have found that modifications
of his wave equation would have been necessary. This point is easily illustrated
by studying the Schr\"odinger-Maxwell system for the hydrogen atom,
\begin{eqnarray}\nonumber
&&\!\!\!\mathrm{i}\hbar\frac{\partial\psi(\bm{x},t)}{\partial t}
+\frac{\hbar^2}{2m}\left(\bm{\nabla}+\mathrm{i}\frac{e}{\hbar}\bm{A}(\bm{x},t)\right)^2
\psi(\bm{x},t)
\\ \label{eq:motpsi}
&&+\,\frac{\alpha_S\hbar c}{|\bm{x}|}\psi(\bm{x},t)=0,
\end{eqnarray}
\begin{eqnarray}\nonumber
&&\!\!\!\left(\frac{1}{c^2}
\frac{\partial^2}{\partial t^2}-\Delta\right)
\bm{A}(\bm{x},t)=
-\,\mu_0\,\frac{e\hbar}{2\mathrm{i}m}
\Big(
\psi^+(\bm{x},t)\bm{\nabla}\psi(\bm{x},t)
\\ \nonumber
&&-\,\bm{\nabla}\psi(\bm{x},t)^+\cdot
\psi(\bm{x},t)
+2\mathrm{i}\frac{e}{\hbar}\psi^+(\bm{x},t)
\bm{A}(\bm{x},t)\psi(\bm{x},t)\Big)
\\ \label{eq:motA}
&&-\,\frac{\mu_0 e}{4\pi}\frac{\partial}{\partial t}\int\!d^3\bm{x}'\,
\frac{\bm{x}-\bm{x}'}{|\bm{x}-\bm{x}'|^3}|\psi(\bm{x}',t)|^2.
\end{eqnarray}
This describes the coupling of the electromagnetic potential in Coulomb gauge,
$\bm{\nabla}\cdot\bm{A}(\bm{x},t)=0$, to the wavefunction $\psi(\bm{x},t)$ for
relative motion in the atom. It is a good approximation if the dominant wavelengths 
of the electromagnetic field are large compared to the Bohr radius. The last term 
in Amp\`{e}re's law (\ref{eq:motA}) in Coulomb gauge comes from the longitudinal
component of the electric field.

A severe difficulty with Schr\"odinger's proposal is that the evolution 
equations (\ref{eq:motpsi},\ref{eq:motA}) for the atom-photon system 
have solutions which correspond to excited atomic energy eigenstates 
{\it without} radiating a photon. Absence of radiation from excited states 
on the level of the wave equations (\ref{eq:motpsi},\ref{eq:motA}) can 
be seen e.g. by observing that substitution of the atomic 
eigenstates $\psi_{n,\ell,0}(\bm{x},t)$ into the equations solves these 
equations for $\bm{A}(\bm{x},t)=0$. For $m_\ell\neq 0$, 
we note that the excited energy eigenstates $\psi_{n,\ell,m_\ell}(\bm{x},t)$ 
yield static vector potentials $\bm{A}(\bm{x})$, but no radiation fields.
Indeed, equations (\ref{eq:motpsi},\ref{eq:motA})
cannot describe emission of classical radiation under any circumstances, since
a Poynting vector for radiation must drop with distance $r$ like $r^{-2}$, whereas
any Poynting vector formed from solutions of (\ref{eq:motA}) with atomic orbitals with
principal quantum numbers $n$ and $n'$ on the right hand side would be exponentially
suppressed at large distance with a factor $\exp[-2(n+n')r/(nn'a_0)]$. 
This implies that the coupling terms of the atom-photon 
system cannot explain spontaneous emission from excited atomic eigenfunctions as a 
consequence of continuous time evolution according to the wave 
equations (\ref{eq:motpsi},\ref{eq:motA}). 
The continually evolving atomic energy eigenfunctions $\psi_{n,\ell,m_\ell}(\bm{x},t)$
would not spontaneously decay through 
emission of electromagnetic radiation at the level of wave equations. 
Even with the coupling terms included, the Schr\"odinger equation and Amp\`{e}re's 
law predict that excited atomic eigenstates would be stable within the 
semi-classical formalism, and no classical radiation could be emitted under
any circumstances. 

Instead, we will see in the following sections that the minimally coupled
Schr\"odinger-Maxwell equations require quantized photon operators to describe
spontaneous emission of radiation. We will also 
see in Sections \ref{sec:mono} and \ref{sec:coherent} that quantum jumps appear as a 
generic property of any electromagnetic interaction of atoms, no matter whether any 
energies are resolved. To elucidate these points, the explicit
or implicit appearance of line spectra will be addressed from the perspective of two 
different optical probes, {\it viz.} monochromatic photon states and coherent photon states, 
to see what the standard calculations of 
photon absorption and emission in quantum mechanics tell us about the dichotomy between 
superpositions of energy eigenstates on the one hand and line spectra on the other hand. 
This will lay the groundwork for explaining how second quantization 
reconciles the deterministic evolution of quantum states with quantum jumps
in Section \ref{sec:conc}.
It will also provide yet another piece of evidence for the importance of field
quantization for interpretations of quantum mechanics, thus supplementing the recent
observations of Myrvold on the relevance of quantum fields for the emergence of 
wavefunctions \cite{myrvold}.

%%%%%%%%%%%%%%%%%%%%%%%%%%%%%%%%%%%%%%%%%%%%%%%%%%%%%%%%%%%%%%%%%%%%%%%%
\section{Absorption and emission of monochromatic photon states}
\label{sec:mono}
%%%%%%%%%%%%%%%%%%%%%%%%%%%%%%%%%%%%%%%%%%%%%%%%%%%%%%%%%%%%%%%%%%%%%%%%

We are interested in photon emission or absorption in the visible and near-ultraviolet
wavelength range, where dipole approximation is excellent and the coupling of 
electromagnetic fields can effectively be described by minimal coupling to the atomic 
wavefunction for relative motion of the electron and proton, rather than coupling to the 
individual constituents of the atom.
The relevant second quantized Hamiltonian for the study of electromagnetic
absorption and emission from a gas of hydrogen atoms is therefore
\begin{equation}\label{eq:Hegamma}
H=H_0+H_I+H_{II},
\end{equation}
where 
\begin{eqnarray*}
H_0&=&\int\!d^3\bm{x}\left(
\frac{\hbar^2}{2m}\bm{\nabla}\psi^+(\bm{x})\cdot\!\bm{\nabla}\psi(\bm{x})
+\frac{\epsilon_0}{2}\dot{\bm{A}}^2(\bm{x})
\right.
\\
&&
-\left.\psi^+(\bm{x})\frac{\alpha_S\hbar c}{|\bm{x}|}\psi(\bm{x})
+\frac{1}{2\mu_0}\left(\bm{\nabla}\times\bm{A}(\bm{x})\right)^2\right)
\end{eqnarray*}
is the Hamiltonian of decoupled hydrogen atoms and external photons
in Coulomb gauge. The constant $\alpha_S=e^2/4\pi\epsilon_0\hbar c$ is 
the fine structure constant.
The interaction terms are the minimal coupling terms
in Schr\"odinger field theory,
\begin{eqnarray} \nonumber
H_I&=&\int\!d^3\bm{x}\,
\frac{e\hbar}{2\mathrm{i}m}\bm{A}(\bm{x})\cdot
[\psi^+(\bm{x})\cdot\bm{\nabla}\psi(\bm{x})
\\ \label{eq:H1}
&&
-\,\bm{\nabla}\psi^+(\bm{x})\cdot\psi(\bm{x})],
\end{eqnarray}
\begin{equation}\label{eq:H2}
H_{II}=\int\!d^3\bm{x}\,\frac{e^2}{2m}\psi^+(\bm{x})
\bm{A}^2(\bm{x})\psi(\bm{x}).
\end{equation}

The Hamiltonian (\ref{eq:Hegamma}) is written in terms of the 
Schr\"o\-din\-ger picture field operators $\psi(\bm{x})$, $\bm{A}(\bm{x})$,
and the electromagnetic field is quantized in Coulomb gauge, i.e. the
photon field in the interaction picture is
\begin{eqnarray}\nonumber
\bm{A}(\bm{x},t)&=&\sqrt{\frac{\hbar\mu_0 c}{(2\pi)^3}}
\int\!\frac{d^3\bm{k}}{\sqrt{2k}}\sum_{\alpha=1}^2
\bm{\epsilon}_\alpha(\bm{k})
\\ \nonumber
&&\times\Big(
a_\alpha(\bm{k})
\exp[\mathrm{i}(\bm{k}\cdot\bm{x}-ckt)]
\\ \label{eq:Axt}
&&+a_{\alpha}^+(\bm{k})
\exp[-\mathrm{i}(\bm{k}\cdot\bm{x}-ckt)]\Big),
\end{eqnarray}
with $\bm{k}\cdot\bm{\epsilon}_\alpha(\bm{k})=0$,
$\bm{\epsilon}_\alpha(\bm{k})\cdot\bm{\epsilon}_\beta(\bm{k})=\delta_{\alpha\beta}$,
$[a_{\alpha}(\bm{k}),a_{\beta}(\bm{k}')]=0$, 
$[a_{\alpha}(\bm{k}),a_{\beta}^+(\bm{k}')]=\delta_{\alpha\beta}\delta(\bm{k}-\bm{k}')$,
and $\dot{\bm{A}}(\bm{x})\equiv\dot{\bm{A}}(\bm{x},0)$ in $H_0$.

Spin labels will be suppressed since we did not include the subleading Pauli term 
and the terms (\ref{eq:H1},\ref{eq:H2}) do not induce spin transitions.
We will first discuss emission rates before deriving absorption cross
sections. In agreement with the superposition principle in quantum mechanics,
no presumption will be made that initial or final atomic states should
be energy eigenstates.

To explore the tension between the superposition principle and the 
emergence of line spectra, we have to study the scattering matrix element for
the transition from the initial state (now written in the language of
second quantization and at $t=0$)
\begin{equation}\label{eq:lineinit}
\bm{|}\{C\}\bm{\rangle}=
\sum_{n,\ell,m_\ell}C_{n,\ell,m_\ell}\!\int\!d^3\bm{x}\,\psi^+(\bm{x})\bm{|}0\bm{\rangle}
\Psi_{n,\ell,m_\ell}(\bm{x})
\end{equation}
to the final state with an emitted photon of momentum $\hbar\bm{k}$
and polarization $\bm{\epsilon}_\alpha(\bm{k})$,
\begin{eqnarray}\nonumber
\bm{|}\{C'\};\bm{k},\alpha\bm{\rangle}&=&
\sum_{n,\ell,m_\ell}\!C'_{n,\ell,m_\ell}\!
\int\!d^3\bm{x}\,\psi^+(\bm{x})a^+_\alpha(\bm{k})\bm{|}0\bm{\rangle}
\\ \label{eq:linefinal}
&&\times
\Psi_{n,\ell,m_\ell}(\bm{x}).
\end{eqnarray}
Here boldface Dirac notation $\bm{|}\ldots\bm{\rangle}$ is introduced for 
the Fock space states of the second quantized theory, to tell them apart from
the states of the first quantized theory like (\ref{eq:super1}). 

Photon absorption and emission amplitudes for transitions in the form of 
scattering matrix elements between energy eigenstates are reported in many 
textbooks, see e.g. \cite{dirac2,heitler,merzbacher,ballentine2,louisqm,rdqm}, 
and just considering those scattering matrix elements in isolation can make 
us easily forget that they actually correspond to a superposition of energy 
eigenstates at least in the final state: Even if we presume that the initial 
state is an energy eigenstate of the unperturbed system,
\[
\bm{|}n(t\to-\infty)\bm{\rangle}=\lim_{t\to-\infty}\exp(-\mathrm{i}H_0t/\hbar)\bm{|}n\bm{\rangle},
\]
the scattering matrix
\begin{eqnarray}\nonumber
S_{n'n}&=&\lim_{t\to\infty,t'\to-\infty}\bm{\langle} n'(t)\bm{|}U(t,t')\bm{|}n(t')\bm{\rangle}
\\  \label{eq:smatrix1}
&=&\bm{\langle} n'\bm{|}U_D(\infty,-\infty)\bm{|}n\bm{\rangle}
\end{eqnarray}
with the interaction picture (or Dirac picture) time evolution operator
\begin{eqnarray}\nonumber 
U_D(t,t')&=&
\exp(\mathrm{i}H_0t/\hbar)
\exp[-\mathrm{i}H(t-t')/\hbar]
\\ \label{eq:UD}
&&\times
\exp(-\mathrm{i}H_0t'/\hbar)
\end{eqnarray}
always produces the final state as a superposition of eigenstates,
\begin{equation}\label{eq:finalstate}
\bm{|}\psi(t\to\infty)\bm{\rangle}=\sum_{n'}\bm{|}n'(t\to\infty)\bm{\rangle} S_{n'n},
\end{equation}
i.e. quantum dynamics does not {\it seem} to describe photon emission or 
absorption as quantum jumps between Bohr levels. However, 
please note that the superposition (\ref{eq:finalstate}) necessarily includes 
states from different sectors in Fock space, which implies that
the state upon observation still includes a reduction e.g. to a single atom
state or to a state containing both an atom and a photon. In that 
sense, $\bm{|}\psi(t\to\infty)\bm{\rangle}$ in (\ref{eq:finalstate}) should be 
considered as an epistemic state, which reduces to an ontic component 
upon observation.

It is noteworthy that the first two papers on time-dependent
perturbations, {\it viz.} Schr\"odinger's early study of light scattering
in Ref. \cite{erwin} and Dirac's first path breaking paper \cite{dirac1a} 
on the development of the general theory of time-dependent perturbations, 
explicitly use linear superpositions of energy eigenstates both in the 
initial and in the final state (however, without addressing the potential 
contradiction with quantum jumps between energy levels in \cite{dirac1a}). 
Dirac's second paper \cite{dirac1b}, which was partly written in Copenhagen,
assumes energy eigenstates as initial states, but still needs to treat
the final states as superpositions of energy eigenstates in agreement
with (\ref{eq:finalstate}).
Therefore the Schr\"odinger equation was a breakthrough in quantum theory 
by providing a dynamical rationale to Heisenberg's transition matrix 
elements as scattering matrix elements, but it also created a problem
by apparently taking the jumps out of the theory, which on the other 
hand seem inevitable to understand observations.

Viewing (\ref{eq:finalstate}) only as a formal sum over transition 
probability amplitudes would suggest that the Schr\"odinger
equation and its relativistic generalizations are just evolution
equations for mnemonic devices
\[
\bm{|}\psi(t)\bm{\rangle}=\sum_n\bm{|}n\bm{\rangle}\bm{\langle} n\bm{|}\psi(t)\bm{\rangle}
\]
which only remind us that the probability to encounter the
eigenstate $\bm{|}n\bm{\rangle}$ at time $t$ is $|\bm{\langle} n\bm{|}\psi(t)\bm{\rangle}|^2$.
Thinking of $\bm{|}\psi(t)\bm{\rangle}$ only as a mnemonic 
device for evolution of probability amplitudes makes the whole 
conundrum of the ``dynamics of collapse of a state'' redundant. This is the 
information explanation (or epistemic view) of quantum 
states \cite{fuchs,qb1,ferrero,spekkens,qb2}, and it holds the promise to 
avoid any need for hidden dynamics or hidden variables without invoking 
problematic assumptions about the observer/system interaction
or the quantum/classical divide\footnote{It is difficult to assess from
Bohr's writings on quantum states and observers how much importance he 
would have assigned
to purely epistemic interpretations. This question is beautifully 
revisited in a recent investigation by Zinkernagel \cite{BohrZ}.}.
Rather than being a dynamical variable attributed to the system in the 
same sense as e.g. location $\langle\bm{x}\rangle(t)$, the 
state $\bm{|}\psi(t)\bm{\rangle}$ summarizes our knowledge about the system, without 
any connotation that there is any deficiency in the sense that there should 
be more to know. As already emphasized before,
this approach has been challenged with numerous counterarguments
\cite{realpsia,realpsi0,realpsi1,realpsi2,realpsi3,realpsi4,realpsi5,realpsi6},
but it offers particularly appealing perspectives on the problem of line spectra,
and we will continue to use the epistemic interpretation for illustrative
purposes. Indeed, both ontic and epistemic views of quantum states have
so many respective advantages that the question may well be: {\it If} these
opposed interpretations can be reconciled in some way, how much of
each interpretation can and must be maintained?

If we suppose that the state $\bm{|}\psi(t)\bm{\rangle}$ just represents our best 
possible knowledge about a quantum system, can we do without quantum jumps to 
explain line spectra? The explicit forms of the scattering matrix elements 
between superpositions of atomic energy eigenstates with different kinds of
photon states in the initial or final state will help us to shed some light 
on this question.

The scattering matrix between eigenstates of $H_0$ to any order
(see e.g. Section 13.7 in Ref. \cite{rdqm}, here $V=H_I+H_{II}$),
\begin{eqnarray}\nonumber
S_{\!fi}&=&\delta_{fi}-2\pi\mathrm{i}\delta(E_f-E_i)V_{fi}
-2\pi\mathrm{i}\delta(E_f-E_i)
\\ \nonumber
&&\times\sum_{n=2}^\infty\sum_{j_1,\ldots j_{n-1}}
V_{fj_1}V_{j_1 j_2}\ldots V_{j_{n-2}j_{n-1}}V_{j_{n-1}i}
\\ \nonumber
&&\times
\left[(E_i-E_{j_1}+\mathrm{i}\epsilon)(E_i-E_{j_2}+\mathrm{i}\epsilon)\ldots
\right. 
\\ \label{eq:Snthorder}
&&\times\left.
(E_i-E_{j_{n-2}}+\mathrm{i}\epsilon)(E_i-E_{j_{n-1}}+\mathrm{i}\epsilon)\right]^{-1},
\end{eqnarray}
implies that superposition of unperturbed states will always yield sums over energy 
preserving $\delta$ functions between unperturbed energy eigenstates (see the 
appendix for an explanation of the emergence of energy preserving $\delta$ functions
in scattering matrix elements). This implies 
that the scattering matrix for bound atomic states will produce line spectra for 
single photons at every order of perturbation theory. We will exlicitly demonstrate 
the emergence of line spectra from superpositions for two different sets of 
observations in first order perturbation theory.

The differential photon emission amplitude for the 
states (\ref{eq:lineinit},\ref{eq:linefinal}) is in first order 
\begin{eqnarray}\nonumber
S_{\!fi}&=&-\,\frac{\mathrm{i}}{\hbar}\int_{-\infty}^\infty\!dt\,\bm{\langle}
\{C'\};\bm{k},\alpha\bm{|}\exp(\mathrm{i}H_0t/\hbar)
H_I
\\ \label{eq:line1}
&&\times\exp(-\mathrm{i}H_0t/\hbar)\bm{|}\{C\}\bm{\rangle}.
\end{eqnarray}
In agreement with equations (\ref{eq:smatrix1},\ref{eq:finalstate})
and generally accepted conventions of applied quantum mechanics,
this is a probability amplitude for the transition from the initial
atomic state $\bm{|}\{C\}\bm{\rangle}$ (\ref{eq:lineinit}) to the final 
state $\bm{|}\{C'\};\bm{k},\alpha\bm{\rangle}$ (\ref{eq:linefinal})
containing an atomic state with expansion coefficients $C'_{n,\ell,m_\ell}$ 
and a photon state with momentum $\hbar\bm{k}$ and 
polarization $\bm{\epsilon}_\alpha(\bm{k})$. 

Evaluation of (\ref{eq:line1}) in dipole approximation yields the superposition 
of the familiar scattering matrix elements for transitions between atomic 
energy eigenstates,
\begin{eqnarray}\nonumber
S_{\!fi}&=&-\,c\sqrt{\alpha_S k}
\sum_{n',n,\ell,m_\ell}\sum_{\Delta\ell=\pm 1}\sum_{\Delta m_\ell=0,\pm 1}
C'^+_{n',\ell+\Delta\ell,m_\ell+\Delta m_\ell}
\\ \nonumber
&&\times
\langle n',\ell+\Delta\ell,m_\ell+\Delta m_\ell|
\bm{\epsilon}_\alpha(\bm{k})\cdot\mathrm{\bf x}|n,\ell,m_\ell\rangle
\\ \label{eq:line2}
&&\times
C_{n,\ell,m_\ell}\delta(\omega_{n',\ell+\Delta\ell;n,\ell}+ck),
\end{eqnarray}
where $\omega_{n',\ell+\Delta\ell;n,\ell}\equiv\omega_{n',\ell+\Delta\ell}-\omega_{n,\ell}$
and dipole selection rules resulting from the matrix element are 
already taken into account. Evaluation of all the quantized field operators
has transformed the second quantized matrix element in (\ref{eq:line1}) 
into a remnant combination of first quantized matrix elements between 
atomic wavefunctions,
\[
\langle n',\ell+\Delta\ell,m_\ell+\Delta m_\ell|
\bm{\epsilon}_\alpha(\bm{k})\cdot\mathrm{\bf x}|n,\ell,m_\ell\rangle
\]
\[
=\int\!d^3\bm{x}\,\Psi^+_{n',\ell+\Delta\ell,m_\ell+\Delta m_\ell}(\bm{x})
\bm{\epsilon}_\alpha(\bm{k})\cdot\bm{x}\Psi_{n,\ell,m_\ell}(\bm{x}),
\]
where we use upright notation $\mathrm{\bf x}$ for the
operator for particle location in quantum mechanics.

%%hier!! explain eq. line3. 
The scattering matrix element (\ref{eq:line2}) yields the differential 
emission rate\footnote{See e.g. \cite{heitler,rdqm} for an explanation of the
relation between scattering matrix elements and emission rates.
The time $\Delta t$ is the observation time window, which is finally taken
as $\Delta t\to\infty$, see the appendix.
Equation (\ref{eq:line3}) gives the differential emission rate in the sense
that the number of photons with polarization $\bm{\epsilon}_\alpha(\bm{k})$,
emitted by $n(\{C\})$ atoms in the 
state $\bm{|}\{C\}\bm{\rangle}$ into a solid angle $\Omega$ and with wavenumbers
$k_1\le k\le k_2$, due to the atomic
transition into the final state $\bm{|}\{C'\}\bm{\rangle}$,
is $n(\{C\})\int_{\Omega}d\Omega\int_{k_1}^{k_2}dk d\Gamma^{(\alpha)}(\bm{k})/d\Omega dk$.
}
$d\Gamma^{(\alpha)}(\bm{k})/d\Omega dk=k^2|S_{\!fi}|^2/\Delta t$ 
in the form
\begin{eqnarray}\nonumber
\frac{d\Gamma^{(\alpha)}(\bm{k})}{d\Omega dk}
&=&\frac{\alpha_S c^2}{2\pi}k^3\sum_{n',n,\ell}\sum_{\Delta\ell=\pm 1}
\Bigg|
\sum_{m_\ell}\sum_{\Delta m_\ell=0,\pm 1}
C_{n,\ell,m_\ell}
\\ \nonumber
&&\times 
C'^+_{n',\ell+\Delta\ell,m_\ell+\Delta m_\ell}
\langle n',\ell+\Delta\ell,m_\ell+\Delta m_\ell|
\\ \nonumber
&&\times 
\bm{\epsilon}_\alpha(\bm{k})\cdot\mathrm{\bf x}|n,\ell,m_\ell\rangle\Bigg|^2
\\ \label{eq:line3}
&&\times
\hbar\delta(E_{n,\ell}-E_{n',\ell+\Delta\ell}-\hbar ck),
\end{eqnarray}
where it is taken into account that the energy preserving $\delta$ function 
usually prevents additional interference besides the interference terms
which arise from energy degeneracy with respect to $m_\ell$. The emergence
of the energy preserving $\delta$ functions 
in (\ref{eq:Snthorder},\ref{eq:line2},\ref{eq:line3}) 
is briefly outlined in the appendix.

The differential emission rate can be readily integrated with respect to 
photon energy, but the spectrally resolved emission rate is more useful for 
our discussion. The important message in (\ref{eq:line2},\ref{eq:line3}) is 
the appearance of the energy preserving $\delta$ function and the appearance 
of interference terms in (\ref{eq:line3}) from $m_\ell$ degeneracy.

Although our initial and final atomic states are superpositions of
energy eigenstates, standard quantum dynamics without any additional
assumptions implies that the energy resolved spectrum (\ref{eq:line3}) from 
transition between those initial and final states yields the same  
spectral lines that we get from transitions between energy eigenstates.
Can we then conclude that observation of the first Balmer line in a spectrum 
implies that all those photons contributing to that line arise from
actual quantum jumps from an $n=3$ state to an $n'=2$ state?
The derivation of (\ref{eq:line3}) does not help to justify such a
conclusion, since it allows for creation of a contribution to the
Balmer line along with contributions to several other lines. 
We have to invoke energy conservation for the single event that
created the observed photon to conclude that an actual
transition from an $n=3$ state to an $n'=2$ state has taken place to create
each of those photons which contribute to the first Balmer line. From an
epistemic point of view, the ensuing reduction of the quantum state 
is only puzzling if we do not take into account the information aspects of the 
quantum state. Observation of one particular component, {\it viz.} the emitted 
photon, has reduced the wavefunction of the atomic component of the atom-photon 
system to an $n=3$ state before emission and an $n'=2$ state after emission, but 
this is only a reflection of our change of knowledge about the system due to
observation of the photon. It does imply, however, collapse of the atoms into 
energy eigenstates during the interaction, since atomic collision will usually 
perturb the eigenstates before and after the emission, and therefore we cannot 
postulate (contrary to Bohr's orbits) that atoms can only exist in energy eigenstates.
Interpretation of the scattering matrix element (\ref{eq:line2}) in terms
of a collapse of atomic states therefore appears inevitable.
This creates a problem if we associate quantum state collapse with observation, 
as in the traditional Copenhagen interpretation, since it would apparently 
introduce an acausal, retroactive aspect into the collapse. The 
matrix element (\ref{eq:line2}) indicates that collapse is not
a consequence of observation, but rather a hallmark of occurence of
an elementary interaction, {\it viz.} spontaneous emission of a photon.

With respect to the photons, we have decomposed the emitted radiation into 
monochromatic photon states through decomposition in that basis of states, 
because Born's rule implies that this is appropriate for the calculation 
of signals in an experiment 
where we would place a spectral polarimeter in direction $\hat{\bm{k}}$ from 
an emitting gas of atoms or molecules. The particular component that emerges from 
a prism or grating spectrograph in a certain direction consists of monochromatic 
photon states of momentum $\bm{k}$. However, that is not because the 
spectrograph collapsed photons into these states in a dynamical sense. 
Instead it scattered the incoming photons coherently into beams of outgoing photons, 
and the basic quantum electrodynamical processes involved imply that incoming photons 
are virtually absorbed by electrons in the spectrograph and re-emitted as outgoing 
photons. The corresponding reduction of the state only signifies what we know about 
the photons emerging in a particular direction from the spectrograph.
There is a difference here compared with the atomic states, since we have to
infer a collapse for the atomic states in ontic representations to make
sense of the energy condition implied with the scattering matrix element (\ref{eq:line2}), 
whereas the photon is spontaneously emitted in agreement with the energy 
condition and no collapse of any pre-existing photon state is needed.

In agreement with unitarity, the first order scattering matrix element for the 
absorption process $\bm{|}\{C'\};\bm{k},\alpha\bm{\rangle}\to\bm{|}\{C\}\bm{\rangle}$
is the negative complex conjugate of the scattering matrix element for the 
inverse emission process (\ref{eq:line1}). The absorption rate is therefore 
given by the same expression (\ref{eq:line3}) as the corresponding emission rate,
\[
\left.\frac{d\tilde{\Gamma}^{(\alpha)}(\bm{k})}{d\Omega dk}
\right|_{\bm{|}\{C'\};\bm{k},\alpha\bm{\rangle}\to\bm{|}\{C\}\bm{\rangle}}
=\left.\frac{d\Gamma^{(\alpha)}(\bm{k})}{d\Omega dk}
\right|_{\bm{|}\{C\}\bm{\rangle}\to\bm{|}\{C'\};\bm{k},\alpha\bm{\rangle}}.
\]
The differential absorption rate can be converted into an absorption
cross section by dividing with the differential current
density per $\bm{k}$ space volume for photons of momentum $\hbar\bm{k}$,
\begin{equation}
\frac{d\bm{j}(\bm{k})}{k^2d\Omega dk}
=\frac{\bm{E}\times\bm{B}}{\mu_0\hbar ck}
=\frac{c}{(2\pi)^3}\hat{\bm{k}}.
\end{equation}
This yields
\begin{eqnarray} \nonumber
\sigma^{(\alpha)}(\bm{k})&=&\frac{d\tilde{\Gamma}^{(\alpha)}(\bm{k})}{k^2dj(\bm{k})}
=4\pi^2\alpha_S ck
\\ \nonumber
&&\times
\sum_{n',n,\ell}\sum_{\Delta\ell=\pm 1}\Bigg|
\sum_{m_\ell}\sum_{\Delta m_\ell=0,\pm 1}\!
C^+_{n,\ell+\Delta\ell,m_\ell+\Delta m_\ell}
\\ \nonumber
&&\times
C'_{n',\ell,m_\ell}\langle n,\ell+\Delta\ell,m_\ell+\Delta m_\ell|
\bm{\epsilon}_\alpha(\bm{k})\cdot\mathrm{\bf x}
\\ \label{eq:abs3}
&&\times
|n',\ell,m_\ell\rangle\Bigg|^2
\delta(\omega_{n,\ell+\Delta\ell;n',\ell}-ck).
\end{eqnarray}
The sharp absorption lines in (\ref{eq:abs3}) are often approximated 
through Lorentzian profiles in practical applications of absorption
cross sections.
However, the important, although at this stage certainly not unexpected
message from (\ref{eq:abs3}) is that quantum dynamics again synthesizes
the response of the system through superposition of photon absorption
between pairs of atomic energy eigenstates without any presumption that 
absorption actually proceeds through energy eigenstates. As before, we 
have to infer quantum state collapse and the direct involvement of the 
atomic energy eigenstates from energy conservation for single photons.

What we will see in spectral decompositions of emission or absorption
spectra will be emission lines (\ref{eq:line3}) or absorption lines
(\ref{eq:abs3}) with photon energies corresponding exactly to the
transition energies between atomic energy eigenstates, even when we
suppose that the transitions happen between superpositions of energy 
eigenstates. The emission and absorption line spectra are in perfect 
agreement with observations, of course, but the emphasis is on the 
observation that direct quantum evolution on the basis of wave equations 
does not automatically imply direct emission or absorption through atomic 
energy eigenstates to predict the observed line spectra. Deterministic
evolution of quantum states remains mute on that particular aspect.
Within the epistemic view, we could nevertheless infer that emission
or absorption of photons proceeds through the corresponding atomic
energy eigenstates. That only seems to violate causality
through retroactive reduction of the state of the atom,
because changing the state only signifies our change in knowledge.
We also do not have to postulate that the atom reacts to our observation,
or that the photon acts as an observer. However, the photon certainly acts
as a messenger of information, and there is nothing controversial about 
that. Therefore a purely epistemic view of quantum states seems to work 
just fine, but there are caveats: On the one hand it is difficult to conceive 
that a system must not have {\it some} unique and ultimately observable
state at a given time and we would have to assign special 
significance to energy eigenstates.

We will re-examine the emergence of line spectra using other probes to gain
 further insights into the interpretational questions from the point of view 
of applied quantum mechanics.

%%%%%%%%%%%%%%%%%%%%%%%%%%%%%%%%%%%%%%%%%%%%%%%%%%%%%%%%%%%%%%%%%%%%%%%%
\section{Absorption and emission of coherent photon states}
\label{sec:coherent}
%%%%%%%%%%%%%%%%%%%%%%%%%%%%%%%%%%%%%%%%%%%%%%%%%%%%%%%%%%%%%%%%%%%%%%%%

As a complementary approach to the interpretation of observations of
absorption or emission of radiation by atoms or molecules, we will now
assume that the experiment prepares or tests the electric or magnetic fields 
of the radiation. 

The electric and magnetic field operators
\[
\bm{E}(\bm{x},t)=-\,\partial\bm{A}(\bm{x},t)/\partial t,\quad
\bm{B}(\bm{x},t)=\bm{\nabla}\times\bm{A}(\bm{x},t),
\]
yield expectation values corresponding to a classical electromagnetic wave,
\begin{eqnarray}\nonumber
\bm{\langle}\bm{\zeta}\bm{|}\bm{A}(\bm{x},t)
\bm{|}\bm{\zeta}\bm{\rangle}
&=&\bm{\mathcal{A}}(\bm{x},t)
=\sqrt{\frac{\hbar\mu_0 c^3}{(2\pi)^3}}
\int\frac{d^3\bm{k}}{\sqrt{2k}}
\\ \nonumber
&&\times
\sum_{\alpha=1}^2
\bm{\epsilon}_\alpha(\bm{k})
\Big(\zeta_\alpha(\bm{k})\exp[\mathrm{i}(\bm{k}\cdot\bm{x}-ckt)]
\\ \label{eq:cohA}
&&
+\,\zeta_{\alpha}^+(\bm{k})\exp[-\mathrm{i}(\bm{k}\cdot\bm{x}-ckt)]
\Big),
\end{eqnarray}
\[
\bm{\langle}\bm{\zeta}\bm{|}\bm{E}(\bm{x},t)
\bm{|}\bm{\zeta}\bm{\rangle}
=\bm{\mathcal{E}}(\bm{x},t)
=-\,\partial\bm{\mathcal{A}}(\bm{x},t)/\partial t,
\]
\[
\bm{\langle}\bm{\zeta}\bm{|}\bm{B}(\bm{x},t)
\bm{|}\bm{\zeta}\bm{\rangle}
=\bm{\mathcal{B}}(\bm{x},t)
=\bm{\nabla}\times\bm{\mathcal{A}}(\bm{x},t),
\]
if we use the coherent photon state \cite{glauber}
\[
\bm{|}\bm{\zeta}\bm{\rangle}=\exp\!\left(\int\!d^3\bm{k}
\left[\zeta(\bm{k})\cdot a^+(\bm{k})
-\zeta^+(\bm{k})\cdot a(\bm{k})\right]\right)\!\bm{|}0\bm{\rangle}
\]
with expectation values for photon number, energy and momentum
\[
\langle n\rangle=\int\!d^3\bm{k}\,\left|\zeta(\bm{k})\right|^2,
\quad
\langle H_0\rangle=\int\!d^3\bm{k}\,\hbar ck
\left|\zeta(\bm{k})\right|^2,
\]
\begin{eqnarray*}
\langle\bm{P}\rangle&=&
\langle\int\!d^3\bm{x}\,\epsilon_0\bm{E}(\bm{x},t)
\times\bm{B}(\bm{x},t)\rangle
\\
&=&\int\!d^3\bm{k}\,\hbar\bm{k}
\left|\zeta(\bm{k})\right|^2.
\end{eqnarray*}
Here the definitions $\zeta(\bm{k})\cdot a^+(\bm{k})
=\sum_{\alpha=1}^2\zeta_\alpha(\bm{k})a^+_\alpha(\bm{k})$,
$\left|\zeta(\bm{k})\right|^2
=\sum_{\alpha=1}^2\zeta^+_\alpha(\bm{k})\zeta_\alpha(\bm{k})$
were used.

Coherent states are often associated with large photon numbers,
when their relative quantum uncertainties  
$\Delta n/\langle n\rangle=\langle n\rangle^{-1/2}$ and
\begin{eqnarray*}
\Delta E/\langle H_0\rangle&=&\Delta P/\langle|\bm{P}|\rangle
\\
&=&\left(\int\!d^3\bm{k}\,k^2
\left|\zeta(\bm{k})\right|^2\right)^{1/2}\!\left/
\left(\int\!d^3\bm{k}\,k
\left|\zeta(\bm{k})\right|^2\right)\right.
\end{eqnarray*}
are small.
However, no principle of quantum mechanics prevents formation of coherent
states for small photon numbers or interaction of coherent photon
states with single atoms or molecules.

The first order scattering matrix element for emission of a coherent photon 
state is again the negative complex conjugate of the 
scattering matrix element for absorption,
\begin{eqnarray}\nonumber
S_{\!fi}&\equiv& S_{\{C'\};\bm{\zeta}|\{C\}}=-\,S^+_{\{C\}|\{C'\};\bm{\zeta}}
\\ \nonumber
&=&-\,\frac{\mathrm{i}}{\hbar}\int_{-\infty}^\infty\!dt\,\bm{\langle}
\{C'\};\bm{\zeta}\bm{|}
\exp(\mathrm{i}H_0t/\hbar) H_I
\\ \label{eq:linec1}
&&\times\exp(-\mathrm{i}H_0t/\hbar)
\bm{|}\{C\}\bm{\rangle},
\end{eqnarray}
and evaluation in dipole approximation yields 
\begin{eqnarray}\nonumber
S_{\!fi}&=&-\Bigg[\sum_{n',n,\ell,m_\ell}\sum_{\Delta\ell=\pm 1}\sum_{\Delta m_\ell=0,\pm 1}
\sum_\alpha\int\!d^2\Omega_{\bm{k}} \sqrt{\alpha_S k^5}
\\ \nonumber
&&\times
\zeta^+_\alpha(\bm{k})
\exp\!\left(-\frac{|\zeta(\bm{k})|^2}{2}\right)\!
\\ \nonumber
&&\times
\langle n',\ell+\Delta\ell,m_\ell+\Delta m_\ell|
\bm{\epsilon}_\alpha(\bm{k})\cdot\mathrm{\bf x}|n,\ell,m_\ell\rangle
\\ \label{eq:linec3}
&&\times
C'^+_{n',\ell+\Delta\ell,m_\ell+\Delta m_\ell}C_{n,\ell,m_\ell}\Bigg]_{k=\Delta\omega/c},
\end{eqnarray}
where the frequency in the condition on the bracket is the
transition frequency, $k=\omega_{n,\ell;n',\ell+\Delta\ell}/c$.

However, the calculation of the spectral photon emission rate
(\ref{eq:line3}) or the absorption cross section (\ref{eq:abs3})
is based on the completeness relation
\begin{eqnarray*}
&&\!\!\!\sum_n\frac{1}{n!}
\int\!d^3\bm{k}_1\ldots\int\!d^3\bm{k}_n
\sum_{\alpha_1,\ldots\alpha_n}a_{\alpha_1}^+(\bm{k}_1)\ldots a_{\alpha_n}^+(\bm{k}_n)
\\
&&\times\bm{|}0\bm{\rangle}
\bm{\langle} 0\bm{|} a_{\alpha_n}(\bm{k}_n)\ldots a_{\alpha_1}(\bm{k}_1)
=1
\end{eqnarray*}
in the photon sector of Fock space. Therefore the calculation of probabilities
for emission or absorption of coherent states requires the use of a cavity 
approximation since the classical decomposition of the identity through coherent 
states of harmonic oscillators can be formulated for coherent photon states
like $\bm{|}\bm{\zeta}\bm{\rangle}$ only in the form
\begin{eqnarray}\nonumber
&&\!\!\!\prod_{\bm{k},\alpha}\int\frac{d\Re\zeta_\alpha(\bm{k})
d\Im\zeta_\alpha(\bm{k})}{\pi}
\\ \nonumber
&&\times\exp\!\left(\zeta_\alpha(\bm{k})a_{\alpha}^+(\bm{k})
-\zeta_{\alpha}^+(\bm{k})a_\alpha(\bm{k})\right)
\bm{|}0\bm{\rangle}
\\ \label{eq:complete2}
&&\times
\bm{\langle} 0\bm{|}\exp\!\left(\zeta^+_\alpha(\bm{k})a_{\alpha}(\bm{k})
-\zeta_{\alpha}(\bm{k})a^+_\alpha(\bm{k})\right)=1.
\end{eqnarray}
The resulting discretization of momentum space amounts to the
substitution 
\[
\frac{1}{(2\pi)^{3/2}}\int\!d^3\bm{k}\to\frac{1}{\sqrt{V}}\sum_{\bm{k}}
\]
in the mode expansion of the vector potential,
and the amplitudes $\zeta_\alpha(\bm{k})$ become dimensionless.
The matrix element
\begin{eqnarray}\nonumber
S_{\!fi}&=&-\,\Bigg[\sum_{n',n,\ell,m_\ell}\sum_{\Delta\ell=\pm 1}\sum_{\Delta m_\ell=0,\pm 1}
\sum_{\alpha,\hat{\bm{k}}} \pi e\sqrt{\frac{2\mu_0 ck^5}{\hbar V}}
\\ \nonumber
&&\times
\zeta^+_\alpha(\bm{k})
\exp\!\left(-\frac{|\zeta(\bm{k})|^2}{2}\right)
\\ \nonumber
&&\times
\langle n',\ell+\Delta\ell,m_\ell+\Delta m_\ell|
\bm{\epsilon}_\alpha(\bm{k})\cdot\mathrm{\bf x}|n,\ell,m_\ell\rangle
\\ \label{eq:linec4}
&&\times
C'^+_{n',\ell+\Delta\ell,m_\ell+\Delta m_\ell}C_{n,\ell,m_\ell}\Bigg]_{k=\Delta\omega/c}
\end{eqnarray}
therefore yields the $\mathcal{O}(\alpha_S)$ differential
transition probabilities
\begin{eqnarray}\nonumber
dP_{\bm{|}\{C\}\bm{\rangle}\to\bm{|}\{C'\};\bm{\zeta}\bm{\rangle}}&=&
dP_{\bm{|}\{C'\};\bm{\zeta}\bm{\rangle}\to\bm{|}\{C\}\bm{\rangle}}
\\ \label{eq:Pfi}
&=&\prod_{\bm{k},\alpha}\!\frac{d\Re\zeta_\alpha(\bm{k})
d\Im\zeta_\alpha(\bm{k})}{\pi}\!\left|S_{\!fi}(\bm{\zeta})\right|^2
\end{eqnarray}
between the in and out states.

The scattering matrix elements (\ref{eq:linec4}) contain already a
summation over photon momentum, and therefore $|S_{\!fi}|^2$ does not 
contain any $\delta$ function any more which generates a line spectrum. 
Indeed,  any signal that tests for the electric or magnetic field
of emitted photons from a low density atomic or molecular gas would observe
a response that will not yield a line spectrum but interference of 
photon energies from many different atomic or molecular transitions.
Observation of emitted radiation through induced electric or magnetic
polarization or electromagnetic forces on probes
would apparently constitute such measurements.
This part of the result is in agreement with the Copenhagen interpretation.
It is significant, however, that the scattering matrix elements (\ref{eq:linec4}) 
still synthesize the signals (\ref{eq:Pfi}) from interference of transitions 
between equidistant pairs of stationary 
states, $\hbar ck=E_{n,\ell}-E_{n',\ell+\Delta\ell}$. They only combine those pairs 
differently through summation over the photon polarizations
and summation over the momentum directions $\hat{\bm{k}}$. 
Why do quantum jumps between energy eigenstates play such 
an unseemingly prominent role in dynamical responses of quantum systems if we
do not even care about energy of any component of the observed system
at any time?

The apparent ubiquity of quantum jumps between Bohr levels even when no
component of the system is prepared or observed in an energy eigenstate
seems very surprising from the perspective of the traditional Copenhagen 
interpretation. It also presents a puzzle for purely epistemic interpretations
of quantum states. Why should the dynamical response of atoms
to coherent electromagnetic probes involve quantum jumps between energy
eigenstates if these are not ontological aspects of atoms?

%%%%%%%%%%%%%%%%%%%%%%%%%%%%%%%%%%%%%%%%%%%%%%%%%%%%%%%%%%%%%%%%%%%%%%%%
\section{Conclusions}
\label{sec:conc}
%%%%%%%%%%%%%%%%%%%%%%%%%%%%%%%%%%%%%%%%%%%%%%%%%%%%%%%%%%%%%%%%%%%%%%%%

We can now return to our original question for the emergence of line
spectra when we spectroscopically resolve the emitted or absorbed
radiation from a thin gas of atoms or molecules. What we observe when 
using a spectrograph in direction $\hat{\bm{k}}$ from the sample is a 
projection onto photon momentum eigenstates $\bm{\langle}\bm{k},\alpha\bm{|}$, 
and is therefore described by the spectral emission 
rate $d\Gamma(\bm{k})/d\Omega dk=\sum_\alpha d\Gamma^{(\alpha)}(\bm{k})/d\Omega dk$, 
see equations (\ref{eq:line1}-\ref{eq:line3}). On the other hand, observing 
electric or magnetic polarization effects from the emitted radiation amounts 
to projection onto coherent final state components $\bm{\langle}\bm{\zeta}\bm{|}$
(\ref{eq:linec1}). However, {\it both signals are synthesized from 
interference of transitions between equidistant pairs of stationary 
states of the unperturbed system}, $\hbar ck=E_{n,\ell}-E_{n',\ell+\Delta\ell}$.
Technically this comes about because evaluation of the matrix elements of 
the operator (\ref{eq:UD}) automatically projects out the $\omega_{fi}=0$
component in the Fourier decomposition of signal amplitudes at every level 
of perturbation theory (\ref{eq:Snthorder}), where $\omega_{fi}=(E_f-E_i)/\hbar$ 
is the transition frequency between unperturbed eigenstates of the atom-photon 
system. {\it This applies even if the observation does not imply any energy 
measurement of any component of the system in the initial or final state}. 
This is certainly demonstrated by equations (\ref{eq:linec1}-\ref{eq:Pfi}).
These observations cast doubts on any notion of observer induced ontological
collapse of wavefunctions of atoms or photons.
On the other hand, the ubiquity of quantum jumps between energy eigenstates 
in optical experiments, even when we do not observe any energy of any 
component of the system, casts doubts on purely epistemic interpretations.
Ontological quantum jumps between atomic energy eigenstates, accompanied
by observer-independent collapse of atomic states, appears as the most
credible interpretation of equations (\ref{eq:linec1}-\ref{eq:Pfi}).

The sharp jump condition (\ref{eq:jump1}) follows only for very long time 
evolution from the initial unperturbed state to the final state, and one might 
suspect that jumps are an artifact of asymptotic behavior for large times. However, 
the scattering matrix for finite times\footnote{The scattering matrix for finite 
time was already implicit in the early papers on time-dependent perturbation theory 
by Dirac \cite{dirac1a,dirac1b} and by Born and Heisenberg (see \cite{BV,BC}), 
although the elegant formulation in terms of time evolution operators was given 
only much later by Dyson \cite{dyson}.} 
maps expansion coefficients of the quantum state at any initial time $t'$ to 
expansion coefficients at any final time $t$,
\begin{equation}\label{eq:psievolve}
\bm{|}\psi(t)\bm{\rangle}=\sum_{n,n'}\bm{|}n'(t)\bm{\rangle} 
S_{n'n}(t,t')\bm{\langle} n(t')\bm{|}\psi(t')\bm{\rangle},
\end{equation}
and this will still yield quantum jumps, albeit with a finite energy uncertainty
of order $\hbar/(t-t')$. Note that the derivation of the scattering matrix 
element (\ref{eq:line2}) only involves the matrix formulation of standard quantum 
mechanical time evolution without any proposition about relative distances between 
components. Neither the completeness relations for eigenstates of $H_0$ nor the 
identification of the time evolution operator $U(t,t')$ nor the evolution equation
(\ref{eq:psievolve}) require any assumption about asymptotics. 
Whether or not the freely evolving states $\bm{|}n(t)\bm{\rangle}$ are a good description 
of the system at time $t$ does not matter for the fact that the complete set of 
scattering matrix elements $S_{n'n}(t,t')$ provides a complete description of 
evolution from time $t'$ to time $t$ as stated in equation (\ref{eq:psievolve}).
 Furthermore, it appears most likely that experimentalists will soon devise single 
photon probing \cite{1pem1,1pem2,1pem3,1pspec1,1pspec2} of atoms over nanoscale 
distances and with sub-femtosecond time resolution \cite{atto1,atto2}, and the 
results will still be described by the scattering matrix element (\ref{eq:line2}) 
which automatically yields the jump equation (\ref{eq:jump1}), albeit with 
a finite energy uncertainty.

The results from Section \ref{sec:erwin} reaffirm that the established framework 
of photon quantization appears as the only viable option to describe photon 
emission and absorption by atoms. The consequences from the results of 
Sections \ref{sec:mono} and \ref{sec:coherent} are then best stated in the 
reformulation of (\ref{eq:psievolve}) in terms of interaction picture 
states, $\bm{|}\psi_D(t)\bm{\rangle}=\exp(\mathrm{i}H_0t/\hbar)\bm{|}\psi(t)\bm{\rangle}$,
\begin{equation}\label{eq:psievolveD}
\bm{|}\psi_D(t)\bm{\rangle}=\sum_{n,n'}\bm{|}n\bm{\rangle} S_{n,n'}(t,t')
\bm{\langle} n'\bm{|}\psi_D(t')\bm{\rangle},
\end{equation}
\[
S_{n'n}(t,t')=\bm{\langle} n'\bm{|}U_D(t,t')\bm{|}n\bm{\rangle}.
\]
Every power of the coupling constant $e$ in the expansion of the time evolution 
operator (\ref{eq:UD}),
\begin{eqnarray}\nonumber
U_D(t,t')&=&\mathrm{T}\exp\Bigg(-\frac{\mathrm{i}}{\hbar}\int_{t'}^t\!d\tau\,
\exp(\mathrm{i}H_0\tau/\hbar)
\\ \label{eq:UD2}
&&\times
(H_I+H_{II})\exp(-\mathrm{i}H_0\tau/\hbar)\Bigg)
\end{eqnarray}
corresponds to a change of the number of photons in $\bm{|}\psi_D(t)\bm{\rangle}$ by 
one unit. This produces $\bm{|}\psi_D(t)\bm{\rangle}$ as a sum over all sectors of 
the Fock space of photons which comply with energy conservation between the initial 
and final states, even if $\bm{|}\psi_D(t')\bm{\rangle}$ was an atomic state. The 
creation or annihilation of each photon constitutes a quantum jump, but the 
probability amplitude for each of these jumps evolves continuously with time.
The resolution of the dichotomy between continuous evolution of quantum states
on the one hand and quantum jumps on the other hand then calls for a mixed
interpretation: Quantum states evolving only within one sector of Fock space
reflect ontological properties of the quantum system, but sums involving different 
sectors of Fock space are epistemic. This explains why scattering matrices of the 
second quantized theory evolve continuously with time, and yet they are transition 
amplitudes for quantum jumps. It also avoids the pitfall of Schr\"odinger's cat:
The sum over contributions from two sectors of Fock space
\begin{eqnarray}\nonumber
\bm{|}\psi(t)\bm{\rangle}&=&A(t)\bm{|}1,0,0;\bm{k},\alpha\bm{\rangle}
\exp[-\mathrm{i}(E_{1}t/\hbar)-\mathrm{i}ckt]
\\ \label{eq:psi2}
&&+\,B(t)\bm{|}2,1,0\bm{\rangle}\exp(-\mathrm{i}E_{2,1}t/\hbar)
\end{eqnarray}
with $|A(t)|^2+|B(t)|^2=1$ is an epistemic sum over ontic components, and therefore
{\it it does not describe a situation where there is and is not a photon in the system}.
In the absence of a decisive observation, the state (\ref{eq:psi2}) only reflects 
our best possible guess regarding the question whether a photon has been emitted
(if we start with the 2p state) or absorbed (if we start with the 1s state). Note 
also that the difference in interpretation of quantum states within a single sector
of Fock space versus states spanning sums over several sectors of Fock space does not 
contradict the recent no-go theorems for epistemic interpretations, which were based
on measurements on entangled states. We cannot observe the 
state (\ref{eq:psi2}) {\it per se} because we cannot define a probe which
tests for the simultaneous presence and absence of a photon, and therefore
we also cannot observe the state as a component in entangled measurements. 

Within the standard computational framework of applied quantum mechanics, 
the following picture emerges:
The singular stochastic event of spontaneous emission (creation) or 
absorption (annihilation) of a photon by an electron forms the ontological
basis for quantum jumps. The probability amplitudes for these singular
stochastic events evolve continuously according to
the minimally coupled equations of quantum optics, although the event
itself is not continuous. The continuous evolution of the epistemic 
states (\ref{eq:psievolve}) with the time evolution operator
$U(t,t')$ is unitary due to $U^+(t,t')=U(t',t)=U^{-1}(t,t')$, 
but the final step of reduction upon observation
will not be unitary in the mathematical sense since the reduction of the
state is not invertible: An atom in a 1s state accompanied by a 
Lyman $\alpha$ photon can arise from spontaneous decay from an infinite 
multitude of 2p states,
\begin{eqnarray} \nonumber
\bm{|}2\mathrm{p}(t)\bm{\rangle}&=&\sum_{m_\ell=-1}^1C_{m_\ell}
\int\!d^3\bm{x}\,\psi^+(\bm{x})\bm{|}0\bm{\rangle}
\Psi_{2,1,m_\ell}(\bm{x})
\\ \label{eq:2pt}
&&\times\exp(-\mathrm{i}E_{2,1}t/\hbar),
\end{eqnarray}
\[
\sum_{m_\ell=-1}^1\left|C_{m_\ell}\right|^2=1.
\]
Both the 2p state (\ref{eq:2pt}) and the resulting atom-plus-photon state
after spontaneous decay,
\begin{eqnarray} \nonumber
\bm{|}1s;\bm{k}(t)\bm{\rangle}&=&\sum_{\alpha=1}^2B_{\alpha}
\int\!d^3\bm{x}\,\psi^+(\bm{x})a^+_{\alpha}(\bm{k})\bm{|}0\bm{\rangle}
\Psi_{1,0,0}(\bm{x})
\\ \label{eq:1sgammat}
&&\times\exp(-\mathrm{i}E_{1}t/\hbar)\exp(-\mathrm{i}ckt),
\end{eqnarray}
\[
\sum_{\alpha=1}^2\left|B_{\alpha}\right|^2=1,
\]
are ontic states within sectors with defined particle numbers in Fock space. 
The observation in Sections \ref{sec:mono} and \ref{sec:coherent} 
that atoms respond to any kind of optical measurement
through quantum jumps between energy eigenstates, even if we do not observe the energy
of any component of the system during any time of an experiment, certainly indicates
that energy and energy eigenstates are ontological properties of atoms. This
gives credibility to the ontic interpretation of energy eigenstates.

The states (\ref{eq:psievolve}), on the other hand, are generically linear
combinations of states from sectors in Fock space with different particle content.
Since, in 90 years of applied quantum mechanics,
we have not developed or encountered any probes or observations which amount to
collapse into a state spanning different sectors in Fock space, it seems natural
to conclude that the states (\ref{eq:psievolve}) are epistemic sums over
ontic components. The reduction of the state to a component in a single sector
in Fock space upon observation, is also epistemic in the sense that it constitutes 
a leap in our knowledge regarding occurence of emission or absorption of photons. 
However, besides the epistemic leap in knowledge upon observation, there is also 
the underlying ontological quantum jump due to spontaneous emission or absorption 
of a photon. Reduction of the quantum state corresponds to the epistemic leap. The 
ontological jump corresponds to a spontaneous transition e.g. from a single atom
in a 2p state to an atom in a 1s state and a Lyman $\alpha$ photon, without
any connotation that the atom-photon system at any time existed in a
superposition of the initial state and the final state. Please note that
both kinds of discontinuous transitions do not need to appear simultaneously.

 From a different angle, the proposal that states spanning several sectors 
in Fock space are epistemic can also be considered as an ontological 
superselection rule\footnote{Traditional superselection rules forbid
particular linear combinations of single-particle states, e.g. with different 
masses or spins, as these states were not considered meaningful in quantum theory. 
See \cite{earman,ruetsche} for discussions of the history and philosophical aspects 
of these superselection rules. Linear combination of single-particle states 
with different quantum numbers would be
interpreted as spontaneous transition of one particle into another
particle in the scattering matrix framework. 
The modern situation regarding the formulation of traditional 
superselection rules has become complex due to the possibility
of inequivalence of mass and flavour eigenstates in spontaneously broken
gauge theories (flavour is the particle physics quantum number underlying 
nuclear isospin). This leads to oscillations between different flavour
eigenstates. The effect has initially been observed in the form of 
oscillations of hadrons as bound quark states, and more recently
through neutrino oscillations \cite{sk,sno}. Violations of the 
most basic Wigner type superselection rule in terms of oscillations
between states of different spin requires spontaneous breaking
of Lorentz invariance and has not been observed. However, quantum field
theories with broken Lorentz symmetry are an active research field 
in theoretical particle physics, see e.g. \cite{lsb1,lsb2} 
and references there.} 
in the sense that we suppose well-defined particle
content of ontological states at any time, whereas the quantum state 
that we can know and calculate via the scattering matrix only provides
us with probabilities for any particular particle content at any time.

As for the necessary inclusion of Fock space, one might worry that
Haag's theorem \cite{haag1,haag2,fraser,ruetsche} on the non-existence of 
creation operators following the time evolution under full interactions
might be a cause for concern. However, note that the beauty of 
quantum field theory in the interaction picture is that it only 
uses free field operators for all calculations, and therefore also 
only employs a Fock space created by field operators which 
are free\footnote{In popular terms, Haag's theorem is sometimes 
phrased as a statement of non-existence of the interaction picture.
This is not correct. The only actual requirement for the interaction picture
is the existence of quantized free field operators, combined with a prescription
on how to move between free Fock space states through Dirac's version of
Dyson's time evolution operator. The actual implication of Haag's theorem
is that the mapping from the freely evolving interaction picture
operators into the Heisenberg picture operators does not imply that
the Heisenberg picture field operators can be used to generate 
a corresponding Fock space, or as Fraser \cite{fraser} and Ruetsche
\cite{ruetsche} have correctly observed, the Heisenberg picture 
operators cannot be directly associated with particles or quanta.} 
in the sense of following
linear wave equations (this also includes wave equations including
classical potentials). The second quantized interaction terms
enter as polynomials in free field operators into the scattering matrix. 
Indeed, the impossibility to map free Fock space unitarily into
a Fock space of states which individually follow the full time
evolution of an interacting field theory lends further credibility
to the proposal that the sums over states from different sectors of Fock
space, as created by the scattering matrix, are epistemic, while states
within single sectors of Fock space can be ontic.

In summary, both epistemic and ontic properties appear as inherent aspects
of quantum states and their dynamics.
The observations of the present paper indicate that states within sectors of Fock
space with well defined particle numbers are likely ontic, whereas states
like (\ref{eq:psievolve}), which evolve under interactions involving spontaneous
annihilation or creation of particles, generically span several sectors of Fock space 
and should be considered epistemic. 
However, there are certainly many more
physical and philosophical aspects to this conjecture that need to be studied.
Identifying and resolving the boundaries (and epistontic overlaps?) of epistemic 
and ontic domains in Fock space and in quantum dynamics promises to be an 
interesting frontier in quantum foundations and philosophy.

\section*{Appendix: Energy conservation in
decay rates and cross sections}
The emergence of the energy preserving $\delta$ function in scattering 
matrix elements $S_{\! fi}$ like (\ref{eq:Snthorder},\ref{eq:line2}) is a 
consequence of Dirichlet's equation \cite{CH,courant}
\begin{equation}\label{eq:delta1}
\lim_{\tau\to\infty}\int_{-\infty}^\infty\!d\omega'\,
\frac{\sin[(\omega-\omega')\tau]}{\pi(\omega-\omega')}f(\omega')
=f(\omega),
\end{equation}
which holds if the function $f(\omega)$ is continuous. Dirichlet's result 
motivates the notion of Dirac's $\delta$ function
(better addressed as a distribution in the mathematical
literature) in the form
\begin{eqnarray}\label{eq:delta2}
\delta(\omega)
&=&\lim_{\tau\to\infty}\frac{\sin(\omega\tau)}{\pi\omega}
\\ \label{eq:delta2b}
&=&\lim_{\tau\to\infty}\frac{1}{2\pi}\int_{-\tau}^{\tau}\!
dt\,\exp(\mathrm{i}\omega t),
\end{eqnarray}
such that equation (\ref{eq:delta1}) can 
be written as 
\begin{equation}\label{eq:delta3}
\int_{-\infty}^\infty\!d\omega'\,\delta(\omega-\omega')f(\omega')
=f(\omega).
\end{equation}
The energy preserving $\delta$ function appears in scattering matrix 
elements through the representation (\ref{eq:delta2b}).

Proofs of Dirichlet's equation can be found e.g. in Sec. 4.13c
in \cite{courant} or Sec. 2.1 in \cite{rdqm}. Indeed, there are
many generalizations of Dirichlet's result through convolution
integrals with other normalized kernel functions \cite{rdqm}, 
and one of particular relevance for quantum mechanics is
\begin{equation}\label{eq:delta4}
\lim_{\tau\to\infty}\int_{-\infty}^\infty\!d\omega'\,
\frac{\sin^2[(\omega-\omega')\tau]}{\pi(\omega-\omega')^2\tau}f(\omega')
=f(\omega).
\end{equation}
This can be used to justify Fermi's trick to deal with the squares
of $\delta$ functions, wich appear in expressions like $|S_{fi}|^2/\tau$
in the calculations of particle reaction rates.
Particle reaction rates enter into the
calculations of decay rates like (\ref{eq:line3}) and cross sections 
like (\ref{eq:abs3}), and equation (\ref{eq:delta4}) can be used to
show that after division by $\tau$, the square of $\delta$ functions 
expressed in the form (\ref{eq:delta2}) yields again one remaining energy 
preserving $\delta$ function through
\begin{equation}\label{eq:delta5}
\lim_{\tau\to\infty}\frac{\sin^2(\omega\tau)}{\pi^2\omega^2\tau}
=\frac{\delta(\omega)}{\pi}.
\end{equation}
The standard textbook 
version $\delta^2(\omega)/\Delta t\to\delta(\omega)/2\pi$ of Fermi's 
trick follows from (\ref{eq:delta5}) if we take into account that the
observation time in  (\ref{eq:delta2b}) is $\Delta t=2\tau$.

\section*{Acknowledgements}
This work was supported in part by NSERC Canada.
I would also like to thank the anonymous referees for comments and
questions which helped to improve the paper.


\begin{thebibliography}{88}

\bibitem{erwin}
E. Schr\"odinger, Annalen der Physik {\bf 386}, 109 (1926)

\bibitem{fuchs}
C.A. Fuchs, ``Quantum Mechanics as Quantum Information (and only a little more)'',
arXiv:quant-ph/0205039

\bibitem{qb1}
C.M. Caves, C.A. Fuchs, R. Schack, 
Physical Review A {\bf 65}, 022305 (2002)

\bibitem{ferrero}
M. Ferrero, Foundations of Physics {\bf 33}, 665 (2003)

\bibitem{spekkens}
R.W. Spekkens, Physical Review A {\bf 75}, 032110 (2007)

\bibitem{realpsi0}
L. Marchildon, Foundations of Physics {\bf 34}, 1453 (2004);
Foundations of Physics {\bf 45}, 754 (2015)

\bibitem{realpsia}
M. Ferrero, D. Salgado, J.L. S\'anchez-G\'omez, Foundations of Physics {\bf 34}, 1993 (2004)

\bibitem{realpsi1}
M.F. Pusey, J. Barrett, T. Rudolph,
Nature Physics {\bf 8}, 475 (2012)

\bibitem{realpsi2}
L. Hardy, International Journal of Modern Physics B {\bf 27}, 1345012 (2013)

\bibitem{realpsi3}
J. Barrett, E.G. Cavalcanti, R. Lal, O.J.E. Maroney,
Physical Review Letters {\bf 112}, 250403 (2014)

\bibitem{realpsi4}
M.S. Leifer, Physical Review Letters {\bf 112}, 160404 (2014)

\bibitem{realpsi5}
C. Branciard, Physical Review Letters {\bf 113}, 020409 (2014)

\bibitem{realpsi6}
M. Ringbauer, B. Duffus, C. Branciard, E.G. Cavalcanti,	A.G. White, A. Fedrizzi,
Nature Physics {\bf 11}, 249 (2015)

\bibitem{qb2}
C.A. Fuchs, N.D. Mermin, R. Schack, American Journal of Physics {\bf 82}, 749 (2014)

\bibitem{ballentine1}
L.E. Ballentine, Reviews of Modern Physics {\bf 42}, 358 (1970)

\bibitem{ballentine2}
L.E. Ballentine: {\it Quantum Mechanics: A Modern Development}.
World Scientific, Singapore (1998)

\bibitem{1pem1} 
P. Michler, A. Kiraz, C. Becher, W.V. Schoenfeld, P.M. Petroff, L. Zhang,
E. Hu, A. Imamoglu, Science {\bf 290}, 2282 (2000)

\bibitem{1pem2} 
A.I. Lvovsky, H. Hansen, T. Aichele, O. Benson, J. Mlynek, S. Schiller,
Physical Review Letters {\bf 87}, 050402 (2001)

\bibitem{1pem3}
P. Ester, L. Lackmann, S. Michaelis de Vasconcellos, M.C. H\"ubner, A. Zrenner,
M. Bichler, Applied Physics Letters {\bf 91}, 111110 (2007)

\bibitem{atto1}
P.B. Corkum and F. Krausz, Nature Physics {\bf 3}, 381 (2007)

\bibitem{atto2}
M.Th. Hassan, T.T. Luu, A. Moulet, O. Raskazovskaya, P. Zhokhov, M. Garg,
N. Karpowicz, A.M. Zheltikov, V. Pervak, F. Krausz, E. Goulielmakis,
Nature {\bf 530}, 66 (2016)

\bibitem{atom1}
M. Uiberacker, Th. Uphues, M. Schultze, A.J. Verhoef, V. Yakovlev, M.F. Kling, 
J. Rauschenberger, N.M. Kabachnik, H. Schr\"oder, M. Lezius, K.L. Kompa, H.-G. Muller, 
M.J.J. Vrakking, S. Hendel, U. Kleineberg, U. Heinzmann, M. Drescher, F. Krausz,
Nature {\bf 446}, 627 (2007)

\bibitem{atom2}
P. Eckle, A.N. Pfeiffer, C. Cirelli, A. Staudte, R. D\"orner, H.G. Muller, M. B\"uttiker,
U. Keller, Science {\bf 322}, 1525 (2008)

\bibitem{atom3} 
E. Goulielmakis, Z.-H. Loh, A. Wirth, R. Santra, N. Rohringer, V.S. Yakovlev, 
S. Zherebtsov, T. Pfeifer, A.M. Azzeer, M.F. Kling, S.R. Leone, F. Krausz,
Nature {\bf 466}, 739 (2010)

\bibitem{atom4}
H. Niikura, H.J. W\"orner, D.M. Villeneuve, P.B. Corkum,
Physical Review Letters {\bf 107}, 093004 (2011)

\bibitem{atom5}
L. Gallmann, J. Herrmann, R. Locher, M. Sabbar, A. Ludwig, 
M. Lucchini, U. Keller, Molecular Physics {\bf 111}, 2243
(2013)

\bibitem{erwin2}
E. Schr\"odinger, The British Journal for the Philosophy of Science
{\bf 3}, 109 (1952)

\bibitem{erwin3}
E. Schr\"odinger, The British Journal for the Philosophy of Science
{\bf 3}, 233 (1952)

\bibitem{Holland}
P.R. Holland: {\it The Quantum Theory of Motion}.
Cambridge University Press, Cambridge (1993)

\bibitem{perovic}
S. Perovic, 
Studies in History and Philosophy of Modern Physics {\bf 37}, 275 (2006)

\bibitem{vonNeumann}
J. von Neumann: {\it Mathematische Grundlagen der Quantenmechanik},
Springer, Berlin (1932). English translation:
{\it Mathematical Foundations of Quantum Mechanics},
Princeton University Press, Princeton (1955)

\bibitem{dirac1a}
P.A.M. Dirac,
Proceedings of the Royal Society of London A {\bf 112}, 661 (1926)

\bibitem{dirac1b}
P.A.M. Dirac,
Proceedings of the Royal Society of London A {\bf 114}, 243 (1927)

\bibitem{born}
M. Born: {\it Atomic Physics}. 2nd edition, Blackie \& Son, London (1937)

\bibitem{myrvold}
W.C. Myrvold, Synthese {\bf 192}, 3247 (2015)

\bibitem{dirac2}
P.A.M. Dirac: {\it The Principles of Quantum Mechanics}. 4th edition,
Oxford University Press, Oxford (1958)

\bibitem{heitler}
W. Heitler: {\it The Quantum Theory of Radiation}. 3rd edition,
Oxford University Press, Oxford (1954)

\bibitem{merzbacher}
E. Merzbacher: {\it Quantum Mechanics}. 3rd edition, Wiley, New York (1998)

\bibitem{louisqm}
L. Marchildon: {\it Quantum Mechanics: From Basic Principles to Numerical 
Methods and Applications}. Springer, New York (2002)

\bibitem{rdqm}
R. Dick: {\it Advanced Quantum Mechanics: Materials and Photons}.
2nd edition, Springer, New York (2016)

\bibitem{BohrZ}
H. Zinkernagel, Studies in History and Philosophy of Modern Physics {\bf 53}, 9 (2016)

\bibitem{glauber}
R.J. Glauber, Physical Review {\bf 131}, 2766 (1963)

\bibitem{BV}
G. Bacciagaluppi, A. Valentini: {\it Quantum Theory at the Crossroads:
Reconsidering the 1927 Solvay Conference}. Cambridge University Press, 
Cambridge (2009)

\bibitem{BC}
G. Bacciagaluppi, E. Crull, 
Studies in History and Philosophy of Modern Physics {\bf 40}, 374 (2009)

\bibitem{dyson}
F.J. Dyson, Physical Review {\bf 75}, 1736 (1949)

\bibitem{1pspec1}
J. Lavoie, J.M. Donohue, L.G. Wright, A. Fedrizzi,
K.J. Resch, Nature Photonics {\bf 7}, 363 (2013)

\bibitem{1pspec2}
J.S. Wildmann, R. Trotta, J. Mart\'in-S\'anchez, E. Zallo,
M. O'Steen, O.G. Schmidt,  A. Rastelli,
Physical Review B {\bf 92}, 235306 (2015)

\bibitem{earman}
J. Earman, Erkenntnis {\bf 69}, 377 (2008)

\bibitem{ruetsche}
L. Ruetsche: {\it Interpreting Quantum Theories}. Oxford University Press,
Oxford (2011)

\bibitem{sk}
Y. Fukuda {\it et al.} (Super-Kamiokande Collaboration), 
Physical Review Letters {\bf 81}, 1562 (1998)

\bibitem{sno}
Q.R. Ahmad {\it et al.} (SNO Collaboration),
Physical Review Letters {\bf 89}, 011301 (2002)

\bibitem{lsb1}
D. Colladay, V.A. Kostelecky,
Physical Review D {\bf 58}, 116002 (1998)

\bibitem{lsb2}
Q.G. Bailey, V.A. Kostelecky, R. Xui,
Physical Review D {\bf 91}, 022006 (2015)

\bibitem{haag1}
R. Haag, Matematisk-fysiske Meddelelser {\bf 29}(12), (1955)

\bibitem{haag2}
R. Haag: {\it Local Quantum Physics: Fields, Particles, Algebras}.
Springer, Berlin (1992)

\bibitem{fraser}
D. Fraser, 
Studies in History and Philosophy of Modern Physics {\bf 39}, 841 (2008)

\bibitem{CH}
R. Courant, D. Hilbert: {\it Methods of Mathematical Physics}, Volume I.
Interscience, New York (1953) %%p. 78, \S 6

\bibitem{courant}
R. Courant, F. John: {\it Introduction to Calculus and Analysis}, Volume II.
Springer, New York (1989) %%p. 484, Sec. 4.13c

\end{thebibliography}
\end{document}